\documentclass[conference]{IEEEtran}
\IEEEoverridecommandlockouts                          

\usepackage{graphicx} 

\usepackage{cite}
\usepackage{longtable}
\usepackage{csquotes}
\usepackage{makecell} 
\usepackage[hidelinks]{hyperref}
\usepackage[table,xcdraw]{xcolor}
\usepackage{tabularx} 

\title{Where 6G Stands Today: Evolution, Enablers, \\and Research Gaps}

\author{Salma Tika$^{\circ}$, Abdelkrim Haqiq$^{\circ}$, Essaid Sabir$^{\bullet}$, Elmahdi Driouch$^*$  \\
$^{\circ}$Hassan First University of Settat, Faculty of Sciences and Techniques,\\
Computer, Networks, Mobility and Modeling Laboratory: IR2M,26000 -
Settat, Morocco. \\
$^{\bullet}$Department of Science and Technology, TÉLUQ, University of Quebec, Montreal, H2S 3L4, Canada. \\
$^*$Department of Computer Science, University of Quebec at Montreal, Montreal, H2L 2C4, Quebec, Canada.\\
Emails: \{salma.tika.doc, abdelkrim.haqiq\}@uhp.ac.ma, essaid.sabir@teluq.ca, driouch.elmahdi@uqam.ca
}

\IEEEoverridecommandlockouts
\IEEEoverridecommandlockouts\IEEEpubid{\makebox[\columnwidth]{979-8-3315-9878-5/25/\$31.00~\copyright~2025~IEEE } \hspace{\columnsep}\makebox[\columnwidth]{ }}

\begin{document}
\maketitle

\begin{abstract}
As the fifth-generation (5G) mobile communication system continues its global deployment, both industry and academia have started conceptualizing the 6th generation (6G) to address the growing need for a progressively advanced and digital society. Even while 5G offers considerable advancements over LTE, it could struggle to be sufficient to meet all of the requirements, including ultra-high reliability, seamless automation, and ubiquitous coverage.  In response, 6G is supposed to bring out a highly intelligent, automated, and ultra-reliable communication system that can handle a vast number of connected devices. 
This paper offers a comprehensive overview of 6G, beginning with its main stringent requirements while focusing on key enabling technologies such as terahertz (THz) communications, intelligent reflecting surfaces, massive MIMO and AI-driven networking that will shape the 6G networks. Furthermore, the paper lists various 6G applications and usage scenarios that will benefit from these advancements. At the end, we outline the potential challenges that must be addressed to achieve the 6G promises.

\end{abstract}
\section{INTRODUCTION}

The wireless industry has continuously evolved and it is is one of the few industry sectors that have kept a fast-growing trend, with each generation introducing higher frequencies, larger bandwidths, and faster data rates \cite{letaief_roadmap_2019}. Since Marconi’s wireless telegraphy in the 19th century, mobile networks have advanced from 1G’s basic voice services to 5G’s
ultra-high-definition 3D data transmission.  Researchers are
 currently focusing on 6G as 5G deployment expands throughout
the world and is anticipated to be realized by 2030. 
\\
With increasing automation across industries, future networks must support billions of interconnected devices, demanding faster speeds, lower latency, and greater reliability \cite{giordani_toward_2020}.
Wireless communication has evolved significantly over the past few decades, with each new generation introducing technologies that push the boundaries of connectivity. Today, 5G networks are being deployed worldwide, bringing major advancements in speed, reliability, and capacity \cite{alwis_survey_2021}. With peak data rates reaching 10 Gbps, ultra-low latency of just 1 millisecond, and the ability to support a massive number of connected devices, 5G has unlocked a range of innovative applications. From Virtual and Augmented Reality (VR/AR) to autonomous vehicles and the Internet of Things (IoT), 5G has laid the foundation for a more connected and intelligent digital world \cite{giordani_toward_2020}. However, as new technologies emerge and demand even greater performance, it is becoming clear that 5G alone will not be enough to support the next wave of applications \cite{rajoria_brief_2022}.
\\
Future technologies such as Holographic Telepresence, Unmanned Aerial Vehicles (UAVs), Extended Reality (XR), and Industry 5.0 require ultra-high data rates, real-time access to powerful computing resources, and extremely high reliability \cite{rajoria_brief_2022}. Moreover, the growing concept of the Internet of Everything (IoE) aims to connect not just people but also machines, sensors, and intelligent systems on an unprecedented scale far beyond 5G’s capabilities. To meet these increasing demands, researchers and industry are already looking ahead to the sixth generation of wireless communication.
Unlike previous generations, 6G will not just be an upgrade, it will be a transformative shift toward “connected intelligence” \cite{letaief_roadmap_2019}. This means that communication networks will go beyond simply transmitting data by integrating artificial intelligence, edge computing, and advanced sensing technologies to create a seamless, intelligent, and highly efficient digital ecosystem. Emerging technologies such as Terahertz (THz) communication, Non-Orthogonal Multiple Access (NOMA), and Reconfigurable Intelligent Surfaces (RIS) will play a crucial role in enabling these capabilities \cite{alwis_survey_2021}. At the same time, sustainability and energy efficiency will be at the core of 6G development, as the rapid growth in data traffic and device connectivity raises critical challenges for power consumption and environmental impact.
As researchers work to define what 6G will look like, industry and academia are actively shaping the roadmap for its development. The goal is to create a wireless network that is not only faster and more powerful but also smarter, more reliable, and more sustainable, paving the way for a truly intelligent and interconnected future.
\\
This paper is organized as follows: Section \ref{sec:evolution} provides an overview of the evolution and key capabilities envisioned for 6G. Section \ref{sec:enhancements} explores the pathway to 6G through IMT-2030 enhancements and innovations, including both extensions of IMT-2020 and the newly introduced capabilities. Section \ref{sec:technologies} presents the key enablers that will drive 6G, such as millimeter wave and terahertz communications, quantum and optical wireless communications, next-generation multiple access schemes, and the role of AI/ML. Section \ref{sec:open_problems} discusses open problems and insights surrounding 6G deployment, including challenges related to security, spectrum, energy efficiency, and system complexity. Section \ref{sec:conclusion} concludes the paper. 
\section{Toward 6G: Evolution and Capabilities}
\label{sec:evolution}
\vspace{-0.5mm}
The development of wireless communication has been marked by continuous advancements, from Marconi's original telegraphy to today's 5th generation (5G) networks \cite{alwis_survey_2021}, with each generation appearing approximately every 10 years in response to increasing technical and social requirements. Beginning with 1G, mobile networks were born with analog voice; 2G took those analog voice networks and made them digital, adding encrypted voice and small data services \cite{siddiky_comprehensive_2025,agrawal_comparison_2022}. The emergence of 3G made mobile Internet services and multicast video calls available, and 4G/LTE made high-speed broadband access possible, leading to the popularity of smart phones and more abundant content provision. The current 5G systems also support enhanced mobile broadband (eMBB), ultra-reliable low-latency communication (URLLC) and massive machine-type communication (mMTC) for applications such as VR/AR, autonomous systems, and Industry 4.0. Nevertheless, even with these advances, 5G is still inadequate for new application areas with demands for global coverage, ultra-high data rates (in the range of Tbps), latency, and extremely low energy consumption. Nowadays, smart mobile terminals, on-body and in-body sensors, and M2M communication are imposing additional challenges to current system infrastructures. As the Internet of Things evolves into the Internet of Everything with hyper-scale device densities and real-time intelligence, 6G is looked upon as a ground breaking networking paradigm. It seeks to provide intelligent, scalable, and sustainable connectivity to support future services such as holographic telepresence, extended reality and space tourism, while overcoming the constraints of existing systems through disruptive technologies and architectures \cite{wang_road_2023,zhang_6g_2019}.
\begin{figure}
    \centering
    \includegraphics[width=0.7\linewidth]{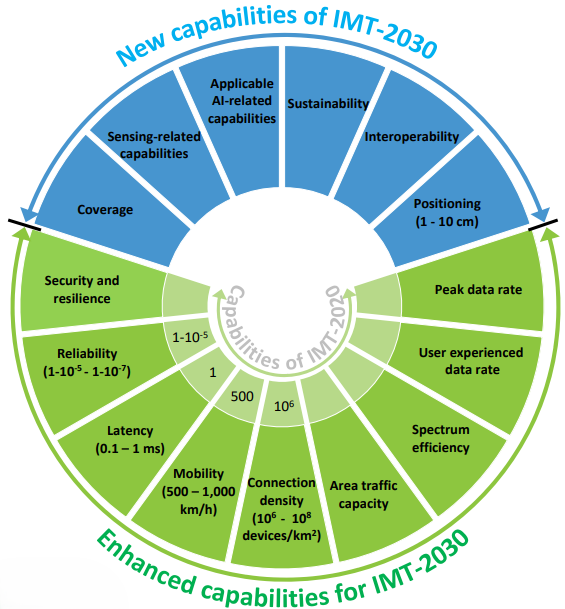}
    \caption{Capabilities of IMT-2020 (5G) and IMT-2030 (6G) \cite{pavithra_nagaraj_demystifying_2024}.}
    \label{fig:capabilities}\vspace{-0.4cm}
\end{figure}
Building on these limitations, 6G is envisioned to revolutionize wireless communication by introducing groundbreaking capabilities that far surpass those of its predecessors. Nine enhanced capabilities of IMT-2030 compared to the capabilities of IMT-2020 \cite{noauthor_imt_nodate} are highlighted in Fig \ref{fig:capabilities}. IMT-2030 (6G) \cite{noauthor_recommendation_nodate}  supports six new capabilities such as coverage, sensing related capabilities, Artificial Intelligence (AI) related capabilities, sustainability, interoperability and positioning.\\
\noindent \textbf{Peak Data Rate:} refers to the maximum data rate achievable under optimal conditions, and 6G aims to reach 1 Tb/s, which is significantly higher than the 0.02 Tb/s offered by 5G \cite{wang_road_2023}.\\
\noindent \textbf{User-Experienced Data Rate:} represents the data rate available to all users within a coverage area, with 6G targeting 10 Gb/s compared to only 0.1 Gb/s in 5G \cite{chen_vision_2020}.\\
\noindent \textbf{Latency:} is the time delay between sending a packet and receiving its acknowledgment. 6G reduces latency to as low as 0.1 ms, ten times lower than the 1 ms latency in 5G \cite{viswanathan_communications_2020}.\\
\noindent \textbf{Spectrum Efficiency:} represents the average data rate per unit of spectrum, with 6G reaching 90 bps/Hz, tripling the 30 bps/Hz efficiency of 5G \cite{chowdhury_6g_2020}.\\
\noindent \textbf{Area Traffic Capacity:} defines the total throughput handled in a given area, with 6G capable of 1 Gbps/m², far exceeding the 10 Mbps/m² capacity of 5G \cite{jiang_road_2021}.\\
\noindent \textbf{Connection Density:} refers to the number of connected devices per unit area, where 6G supports up to $10^8$ devices/km², which is 100 times more than 5G’s $10^6$ devices/km² \cite{wang_road_2023}.\\
\noindent \textbf{Mobility:} refers to the maximum speed at which a connection can be maintained with quality of service, where 6G supports up to 1000 km/h, doubling 5G’s limit of 500 km/h \cite{tataria_6g_2021}.\\
\noindent \textbf{Reliability:} indicates the proportion of successfully delivered packets within a set time, with 6G enhancing this to 99.9999 \%, compared to 99.9\% in 5G \cite{jiang_road_2021}.\\
\noindent \textbf{Security Capacity:} is the rate of secure and reliable data transmission. 6G is expected to deliver much higher security capacity than the relatively low-level granted in 5G \cite{siddiky_comprehensive_2025}.\\
\noindent \textbf{Coverage:} indicates the percentage of an area with network availability, and 6G aims for over 90\% coverage, a substantial improvement from 5G’s 10\% \cite{wang_road_2023}.\\
\noindent \textbf{Sensing-related Capabilities:} is the system’s ability to detect and interpret fine-grained visual or environmental information, where 6G offers millimeter-level accuracy, an improvement over 5G’s one-meter precision \cite{wang_road_2023}.\\
\noindent \textbf{Applicable AI-related capabilities:} refers to the integration of AI into the network for intelligent decision-making, and 6G will feature a high level of intelligence, while 5G remains relatively basic in this regard \cite{wang_road_2023}.\\
\noindent \textbf{Sustainability and Energy Efficiency:} refers to the ability to minimize gas emissions and other environmental impacts throughout the life cycle of a system. It uses multiple services for increased energy efficiency or of optimized use of resources taking place through means such as service life extension, maintenance, repair, remanufacturing, refurbishing, reuse or recycling. Energy efficiency is a measurable indicator that quantifies the system sustainability, as it refers to the number of information bits transmitted per unit of energy \cite{pavithra_nagaraj_demystifying_2024}.\\
\noindent \textbf{Interoperability:} supports communication and cooperation of different components, devices, and networks in the IMT-2030 environment, which is critical to enable different use cases, applications and services \cite{pavithra_nagaraj_demystifying_2024}.\\
\noindent \textbf{Positioning:} refers to the ability to calculate the approximate position of connected devices \cite{pavithra_nagaraj_demystifying_2024}.

\section{PATHWAY TO 6G: IMT 2030 ENHANCEMENTS AND 
INNOVATIONS}
\label{sec:enhancements}

As wireless communication systems evolve from 5G to 6G, the foundational application scenarios introduced in IMT-2020 \cite{noauthor_imt_nodate}—namely enhanced Mobile Broadband (eMBB), ultra-reliable low-latency communication (URLLC), and massive machine-type communication (mMTC)—have been greatly improved and innovated \cite{pavithra_nagaraj_demystifying_2024}. The IMT-2030 \cite{noauthor_recommendation_nodate} framework, proposed for 6G, builds upon these three usage scenarios by introducing evolved versions designed to meet the increasingly complex demands and  to support intelligent and immersive services. In addition to extending the existing use cases, 6G is expected to enable a new generation of transformative applications and capabilities that were beyond the reach of previous systems. Furthermore, 6G  is envisaged to enable new usage scenarios driven by emerging capabilities  such as artificial intelligence and sensing \cite{pavithra_nagaraj_demystifying_2024}, which previous generations were not designed to support. Together, the enhanced and newly introduced scenarios in 6G signal a fundamental shift toward intelligent, immersive, and globally accessible communication networks, setting the stage for the next era of wireless communication.
\begin{figure}
    \centering
    \includegraphics[width=0.8\linewidth]{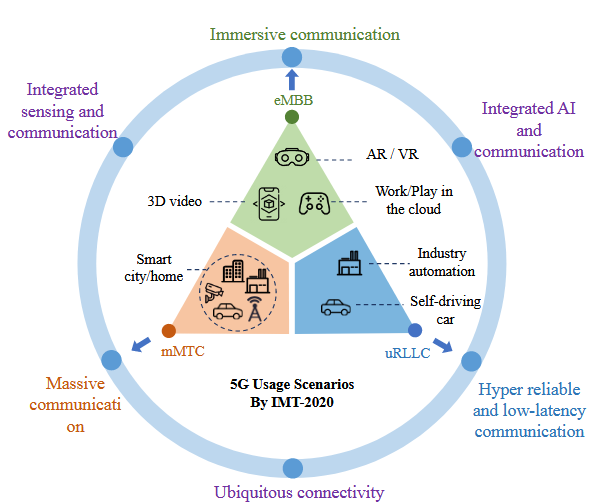}
    \caption{6G Usage Scenarios By IMT-2030 \cite{huang_survey_2019}.}
    \label{fig:usage sc}
\end{figure}
\subsection{6G Enhancements: IMT 2020 Extensions}

\noindent\textbf{Immersive Communication:} As one of the key enhancements of IMT-2030, this usage scenario extends the enhanced Mobile Broadband (eMBB) of IMT-2020 aiming to support deeply interactive, multi-sensory experiences that go far beyond traditional broadband capabilities.  This evolution reflects a shift from the primary 5G goals of  increasing data rates and network capacity which are the central goals of eMBB in 5G, toward enabling deeply immersive applications such as extended reality (XR), holographic telepresence, ultra-HD video streaming, and real-time 3D multi-sensory communication \cite{noauthor_recommendation_nodate}. These use cases demand not only ultra-high data rates but also highly synchronized transmission of video, audio, and environmental data, with some requiring support for low latency and high reliability to ensure seamless interaction with real and virtual environments.  It also addresses new requirements for spectral efficiency and increased system capacity to cope with densely connected devices including hotspots as well as urban and rural areas. Overall, this shift represents a wider evolution from high-speed broadband towards smart and immersive communication services \cite{pavithra_nagaraj_demystifying_2024}.

\noindent\textbf{Hyper Reliable and Low-Latency Communication:} As an evolution of  the Ultra Reliable and Low-Latency Communication usage scenario (URLLC) in IMT-2020, Hyper Reliable and Low-Latency Communication (HRLLC) in IMT-2030  is designed to meet even more stringent requirements for reliability and latency [1].  While 5G URLLC supported  applications that required ultra-reliable and low-latency communication for applications requiring instantaneous response, HRLLC pushes these boundaries further by enabling highly reliable performance.  This enhanced scenario is essential for  time-synchronized operations where communication failure can lead to severe consequences such as in industrial automation, autonomous systems, and healthcare. Use cases include real-time control of robotic systems in manufacturing, remote medical procedures such as telesurgery, and low-latency coordination among connected autonomous vehicles. Moreover, HRLLC supports use cases demanding high connection density and precise positioning, such as smart grid control, environmental monitoring, and intelligent agriculture [1]. These advancements reflect a shift from simply achieving low toward latency enabling a new class of intelligent applications that require instant and reliable data exchange \cite{kaushik_towards_2024, ishteyaq_unleashing_2024}.
\begin{table*}[h]
\caption{6G Usage Scenarios, Key Requirements, and Use Cases.}
\centering
\footnotesize  
\renewcommand{\arraystretch}{1.3}  
\setlength{\tabcolsep}{4pt}  

\resizebox{\textwidth}{!}{  
\begin{tabular}{|>{\columncolor[HTML]{FFFFFF}}p{4cm}|p{5.5cm}|p{5.5cm}|}
\hline
\rowcolor[HTML]{C0C0C0}
\centering \textbf{Usage Scenario} 
&\centering \textbf{Key Requirements}
&   \textbf{ \hspace{2cm} Use Cases}  
\\
\hline

\centering Immersive Communication 
& \makecell[l]{-Ultra-high data rates\\ -Low latency\\ -High reliability\\ -Time-synchronized media\\ -High mobility and capacity}
& \makecell[l]{-Ultra-HD streaming\\ -Metaverse\\ -Mixed/Extended Reality\\ -Holographic communication\\ -Telepresence\\ -Telemedicine} \\
\hline

\centering Hyper Reliable and Low-Latency Communication (HRLLC) 
& \makecell[l]{-Ultra-low latency\\ -Ultra-high reliability\\ -Precise positioning\\ -High connection density\\ -Time-synchronized operation}
& \makecell[l]{-Industrial automation\\ -Remote surgery\\ -Autonomous vehicles\\ -Robotics\\ -Multiplayer VR gaming\\ -Smart grid and agriculture} \\
\hline

\centering Massive Communication 
& \makecell[l]{-High connection density\\ -Low power consumption\\ -Wide coverage\\ -Support for sporadic data\\ -Mobility support}
& \makecell[l]{-Smart cities\\ -Home automation\\ -Environmental/health monitoring\\ -Logistics and supply chains\\ -Smart agriculture} \\
\hline

\centering Ubiquitous Connectivity 
& \makecell[l]{-Affordable universal access\\ -Wide/rural area coverage\\ -Bridging the digital divide}
& \makecell[l]{-Broadband in remote areas\\ -IoT in underserved regions} \\
\hline

\centering AI and Communication (AIAC) 
& \makecell[l]{-High data rate\\ -Low latency and high reliability\\ -Distributed/on-device AI\\ -AI orchestration and inference}
& \makecell[l]{-Autonomous driving\\ -Medical/environmental monitoring\\ -NLP and translation\\ -On-device LLMs\\ -Digital twins} \\
\hline

\centering Integrated Sensing and Communication (ISAC) 
& \makecell[l]{-High-precision positioning\\ -Range/velocity/angle estimation\\ -Sensing, imaging, mapping, detection}
& \makecell[l]{-Gesture/posture recognition\\ -Surveillance/navigation\\ -Fire detection\\ -Traffic optimization\\ -AI/VR/Digital twin sensing} \\
\hline

\end{tabular}
}
\label{tab:6g_scenarios}\vspace{-0.1cm}
\end{table*}
\\
\noindent\textbf{Massive Communication:} Massive Communication in IMT-2030 represents a very advanced version of
mMTC usage scenario in IMT-2020 and further enhances its capabilities to support a significantly higher number of connected devices while also improved scalability and network efficiency \cite{pavithra_nagaraj_demystifying_2024}. While mMTC in 5G primarily focused on connecting a large number of devices for machine-to-machine (M2M) communications, the IMT-2030 framework represents a significant
advancement toward supporting a hyper-connected world comprising a wide range of user applications and device types. This new framework supports adaptive communication across diverse domains. Typical use cases include smart cities, transportation, logistics, healthcare monitoring, environmental sensing, agriculture, and smart grid systems. This usage scenario must support extremely high connection density, ultra-low power consumption, mobility, wide coverage, and scalable network access, alongside high levels of security and reliability. Ultimately, this evolution contributes to
the realization of 6G as a pervasive and intelligent network, capable of interconnecting the digital and physical
worlds \cite{kaushik_towards_2024}.
\subsection{6G Innovations: Newly Added Capabilities}
\noindent\textbf{Ubiquitous Connectivity:} Ubiquitous Connectivity aims to provide affordable and meaningful network access to all users, regardless of their location, including areas that are currently inadequately connected or entirely unconnected, such as low-density or geographically isolated areas. Although 5G marked significant evolution in enhancing coverage and capacity, it remains constrained in high-mobility and hard-to-reach scenarios. in remote, rural, and high-mobility environments.  6G seeks to overcome these gaps by ensuring truly global connectivity across all terrains, including air, sea, and space. This demands the integration  of diverse technologies such as non-terrestrial networks (NTNs) to extend coverage and reconfigurable intelligent surfaces (RIS), which collectively support broader and more adaptive coverage \cite{pavithra_nagaraj_demystifying_2024}.
Typical use cases include, but are not limited to, mobile broadband communication and Internet of Things (IoT) applications \cite{noauthor_recommendation_nodate}.
\\
\noindent\textbf{Artificial Intelligence and Communication (AIAC):} The integration of Artificial Intelligence (AI) into communication systems represents a major shift toward building intelligent, adaptive, and  secure 6G wireless networks. AI contributes significantly by playing a key role in enabling autonomous decision-making, real-time optimization, and predictive resource management across all network layers. This usage scenario also supports high area traffic capacity and user-experienced data rates, as well as low latency and high reliability. Typical applications include assisted automated driving, autonomous collaboration between devices for medical assistance, patient and environmental monitoring applications, digital twin creation and prediction, and
 the offloading of heavy computational operations across devices and networks.
\noindent\textbf{Integrated Sensing and Communication (ISAC):} This use-case marks a significant  advancement in 6G wireless networks by combining precise sensing with advanced communication systems. This integration supports emerging applications   that rely on sensing capabilities. Typical use cases include activity recognition (e.g., gesture or posture detection), movement tracking (e.g., vehicle and pedestrian detection), environmental monitoring (e.g., pollution or fire detection), and the provision of contextual data for AI, XR, and digital twin applications. To meet the demands of these applications, ISAC requires enhanced capabilities including high-precision positioning, range, velocity, and angle estimation. It also enables object localization, imaging, and mapping \cite{wen_survey_2024}. The usage scenarios, along with their corresponding requirements and representative use cases, are summarized in Table \ref{tab:6g_scenarios}.

\section{Key Enabling Technologies of 6G Networks}
\label{sec:technologies}
To support the six usage scenarios of IMT-2030, 6G networks will require a robust set of cutting-edge technologies. These key enablers will address demanding requirements such as ultra-low latency, high reliability, massive connectivity, and energy efficiency. The following highlights key candidate technologies anticipated to play a central role in 6G advancement.

\subsection{Millimeter Wave and Terahertz Communications}
Millimeter wave (mmWave) and Terahertz (THz) communication are becoming increasingly prominent as 5G evolves and 6G takes shape, thanks to their potential to support ultra-fast, high-capacity wireless networks. mmWave was first introduced in 5G, operates in the range 30-300GHz, which provides  significantly more bandwidth compared to traditional bands \cite{pavithra_nagaraj_demystifying_2024}. Despite its limitations—such as high path loss, sensitivity to blockages, and limited diffraction—mmWave remains vital for short-range, high-throughput scenarios. However, to truly unlock data rates in the terabit-per-second range and support emerging applications like holographic calls, extended reality (XR), and real-time sensing, we need to push even further into the spectrum — into the THz range (0.1 to 10 THz),  offering not only extreme data rates but also enabling high-resolution sensing, imaging, and positioning \cite{ishteyaq_unleashing_2024}. At the same time, challenges such as signal blockage, limited coverage distance, and hardware complexity still need to be overcome.  Together, mmWave and THz communications are expected to form the backbone of 6G’s ultra-fast, intelligent, and immersive wireless experiences.
\subsection{Quantum Communications}
As we move beyond 5G, ensuring faster, more secure, and intelligent communication becomes essential. Quantum communication is becoming a promising candidate for addressing future communication requirements. By using principles of quantum mechanics—specifically superposition (particles existing in multiple states), entanglement (correlation across distance), and the uncertainty principle (limits on measurement precision)—to transmit information in a way that makes eavesdropping nearly impossible. Data can be encoded in quantum states e.g., using photons, that can’t be accessed or copied  without detection, enabling security ensured  through principles such as quantum key distribution. In addition to secure transmission, quantum systems can enhance data transmission capacity through the superposition of qubits.
Furthermore, integrating quantum computing with future wireless networks may enhance AI capabilities in 6G, enabling faster and smarter data processing.
 Although still under development, quantum technologies are anticipated to play a foundational role in the future of ultra-secure and intelligent communication infrastructures \cite{zhang_6g_2019}.
\subsection{Optical Wireless Communications (OWCs)}
Optical wireless communications (OWC) have gained increasing attention as an enabling technology for 6G networks,  particularly in scenarios demanding  ultra-high data rates, low latency, high energy efficiency, and enhanced physical layer security. They operate within the optical spectrum range, which includes infrared (IR), visible light (VL), and ultraviolet (UV) bands, providing large unlicensed bandwidth resources. 
Among the key OWC technologies are Visible Light Communication (VLC), Light Fidelity (Li-Fi), and Optical Camera Communication (OCC), each offering unique characteristics and use cases as presented in Table \ref{tab:owc-mini}.
\vspace{-1em}

\begin{table}[h]
\caption{Types of OWC Technologies and Their Use Cases.}
\centering
\scriptsize
\begin{tabular}{|c|c|c|}
\hline
\rowcolor[HTML]{C0C0C0} 
\textbf{Tech} & \textbf{Description} & \textbf{Use Case} \\ \hline
\textbf{VLC} & \makecell{Uses LED light \\ for data transmission} & \makecell{Indoor comm., \\ lighting + internet} \\ \hline
\textbf{Li-Fi} & \makecell{VLC using intensity \\ modulation for high-speed data} & \makecell{Wi-Fi alternative, \\ high-speed access} \\ \hline
\textbf{OCC} & \makecell{LEDs as transmitters, \\ cameras as receivers} & \makecell{Short-range, \\ smartphone-based comm.} \\ \hline
\end{tabular}
\label{tab:owc-mini}
\end{table}
While OWC presents numerous advantages, it also encounters several technical challenges, such as limited coverage, strict line-of-sight (LOS) requirements, inter-cell interference, and mobility management \cite{pennanen_6g_2025}.
\\
Nevertheless, VLC, Li-Fi, and OCC have great potential to be integrated into 6G networks and in particular, they could be used in indoor environments such as hospitals and airports, vehicular networks (V2X), underwater communication, and remote area connectivity.
\subsection{Next Generation Multiple Access (NGMA) Schemes} 
Next-Generation Multiple Access (NGMA) schemes are designed to meet 6G requirements such as massive connectivity, spectral efficiency, and support for diverse use cases. By moving beyond traditional orthogonal access, NGMA techniques such as NOMA, SCMA, D-OMA, and grant-free aim to efficiently serve heterogeneous users and applications across future wireless networks.
\\
Non-Orthogonal Multiple Access (NOMA) is a key multiple access technique for 6G that allows multiple users to share the same time-frequency resources simultaneously, which helps improve spectral efficiency and connectivity. NOMA is mainly categorized into Power-Domain NOMA (PD-NOMA) and Code-Domain NOMA (CD-NOMA). In PD-NOMA, users are distinguished by their power levels, using superposition coding at the transmitter and successive interference cancellation (SIC) at the receiver. In contrast, CD-NOMA uses unique  non-orthogonal spreading codes to separate users. NOMA is especially promising  for massive IoT deployments and grant-free access in 6G, offering support for high device density and low-latency requirements compared to traditional orthogonal multiple access (OMA) schemes \cite{pavithra_nagaraj_demystifying_2024, pennanen_6g_2025, siddiky_comprehensive_2025}.

Sparse Code Multiple Access (SCMA) is a promising code-domain NOMA technique for 6G communications, offering high spectral efficiency and support for massive connectivity. In SCMA, each user is assigned a unique sparse, multidimensional codebook, allowing multiple users to share the same time-frequency resources non-orthogonally. These codebooks enable the spreading of modulated symbols over allocated resources, making simultaneous user detection possible. At the receiver end, the Message Passing Algorithm (MPA) which helps decode and separate all the users’ messages based on their unique codes, which leverages the sparsity of the codebooks to reduce complexity compared to maximum likelihood detectors. SCMA is especially suitable for dense IoT scenarios in 6G, where many devices (like in IoT networks) need to connect at once, offering high spectral efficiency with manageable decoding complexity, making it a strong candidate for next-
generation multiple access in 6G \cite{shah_survey_2021, pavithra_nagaraj_demystifying_2024}. 

Delta-Orthogonal Multiple Access (D-OMA) is a multiple access technique proposed for 6G networks, designed to reduce receiver complexity and power consumption in large-scale in-band NOMA. It allows adjacent NOMA clusters to partially overlap in frequency by a controllable factor ($\delta$), enabling a trade-off between spectral efficiency and receiver complexity. When there is no overlap, D-OMA behaves like traditional power-domain NOMA. Built upon coordinated multipoint (CoMP) transmission, D-OMA thus enhances spectral efficiency by carefully balancing interference and cluster size \cite{akbar_challenges_2025}. Additionally, it enables enhanced security in uplink and downlink communications and supports massive connectivity, making it suitable for 6G applications \cite{shah_survey_2021}.

In ultra-massive IoT implementations, Grant-Free (GF) random access presents a promising solution for fast and efficient communication enabled in 6G. Unlike grant-based schemes, GF approaches enable devices to transmit small packets without prior permission from a base station, greatly improving the latency, signaling overhead, and energy consumption in the process \cite{prasad_tera_toward_2025}. This reduces latency, signaling overhead, and energy consumption. While this approach simplifies access and supports massive connectivity, it also brings challenges as more devices try to access the network simultaneously, issues like collisions and interference can arise. Future 6G systems aim to overcome these challenges through smart collision management and AI-driven traffic prediction, making GF access a practical and scalable choice for next-generation wireless networks. Overall, GF access is expected to play a central role in achieving the speed, scalability, and energy efficiency demanded by 6G networks \cite{pennanen_6g_2025}.
\vspace{-1mm}
\subsection{Reconfigurable Intelligent Surface (RIS)}  
\vspace{-1mm}
Reconfigurable Intelligent Surfaces (RIS), enabled by advances in electromagnetic (EM) metasurfaces and material science, have emerged as a transformative technology for 6G wireless communication systems, aiming to reshape the wireless environment by intelligently controlling the propagation of electromagnetic waves toward intended receivers or away from interference sources, RIS can significantly improve signal-to-noise ratios, extend coverage in challenging environments (e.g., urban blockages or indoor areas), and reduce energy consumption. This makes RIS particularly suitable for scenarios with stringent requirements on energy efficiency, spectral efficiency, and connectivity density. Compared to traditional relays or massive MIMO systems, RIS offers a low-cost, energy-efficient, and flexible solution that can be deployed on various surfaces. On the other hand, integrating RIS into network architectures introduces numerous technical difficulties. These include  low-complexity channel estimation, real-time surface reconfiguration, scalability to large arrays, and efficient coordination with base stations. There is ongoing research utilizing AI algorithms to achieve adaptive and scalable RIS for future applications requiring high density and low latency \cite{pennanen_6g_2025, wang_road_2023, prasad_tera_toward_2025}.

\subsection{Extremely Large Scale and Cell Free Massive MIMO} 
As wireless communication continues its transition toward 6G, MIMO (Multiple-Input Multiple-Output) technologies are undergoing profound evolution to meet the demands of ultra-high capacity, energy efficiency, and intelligent network adaptability. Two of the most groundbreaking advances in this domain are Extremely Large Scale MIMO (XL-MIMO) and Cell-Free Massive MIMO (CF-mMIMO).

XL-MIMO, also referred to as Hyper-MIMO or Ultra-Massive MIMO, pushes the boundaries of wireless communication by deploying an extremely high number of antennas—often in the hundreds or thousands \cite{ishteyaq_unleashing_2024}. This approach unlocks new levels of spectral efficiency, coverage, and energy savings, especially when combined with near-field communications and advanced beamforming techniques. While still under active research, XL-MIMO is expected to play a central role in enabling the high-capacity and high-precision requirements of future 6G networks \cite{pavithra_nagaraj_demystifying_2024,wang_road_2023}. Meanwhile, CF-mMIMO offers a different architectural revolution. Unlike traditional fixed cells where users are served by fixed base stations, CF-mMIMO deploys a vast number of distributed access points (APs) connected to a central processing unit (CPU), collaboratively serving users without rigid cell boundaries. This distributed design mitigates the cell-edge performance degradation seen in conventional networks and enables more uniform quality of service (QoS) across the entire coverage area. This results in shortening link distances, reducing interference, and improving spectral and energy efficiency. A high-level comparison of Massive MIMO and Cell-Free Massive MIMO is provided in table \ref{tab:mimo-comparison}
 for clarity \cite{pennanen_6g_2025, kaushik_towards_2024}. 
\vspace{-3mm}
\begin{table}[htbp]
\caption{Traditional vs Cell-Free Massive MIMO.}
\centering
\scriptsize
\begin{tabular}{|l|l|l|}
\hline
\rowcolor[HTML]{C0C0C0} 
\textbf{Feature} & \textbf{Traditional Massive MIMO} & \textbf{Cell-Free Massive MIMO} \\ \hline
\textbf{Antennas location} & Centralized (at base station) & Distributed (small APs) \\ \hline
\textbf{User served by} & Single base station & \makecell{All nearby APs cooperatively} \\ \hline
\textbf{Cell boundaries} & Fixed & \makecell{No cells — users move freely} \\ \hline
\textbf{Performance} & High at center, weak at edge & Uniform across entire area \\ \hline
\textbf{Handover} & Needed between cells & \makecell{Not needed —  seamless} \\ \hline
\end{tabular}
\label{tab:mimo-comparison}
\end{table}
\vspace{-1em}
\subsection{Energy-Aware Technologies for 6G}
With the growing number of connected devices and increasing data demands, improving energy efficiency is a core priority for 6G. To achieve sustainability without compromising performance, 6G is expected to integrate a range of energy-aware technologies such as green networks, energy harvesting, and AI-based energy management \cite{pennanen_6g_2025, prasad_tera_toward_2025}. Green Networks aim to reduce energy consumption across the network, communication, and device levels. Some of the strategies include powering down underutilized base station components, optimizing resource allocation (e.g., RISs, grant-free access), and adopting energy-aware PHY designs.  On the device level, optimizing transmission power and sleep modes contributes to lower energy use.  AI and ML enhance adaptive, real-time optimization, thereby lowering the overall energy consumption \cite{huang_survey_2019}. Particularly for large scale IoT systems, \textbf{\textit{Energy Harvesting}} (EH) outlines new progress toward low-power and autonomous devices. By capturing energy from ambient sources such as RF signals, solar, or kinetic activity, devices can operate with minimal reliance on traditional power supplies. However, realizing large-scale EH in 6G requires advanced materials, antenna design, and hybrid architectures are also needed to make EH viable for large-scale 6G deployment\cite{ibhaze_brief_2022, akbar_challenges_2025}.

\subsection{AI/ML in 6G}
Like all previous generations, the sixth generation (6G) of wireless communication is expected to improve data rates and reduce latency, but it also projected to radically transform networks into intelligent, autonomous, and adaptive frameworks. Artificial Intelligence (AI) and Machine Learning (ML) will revolutionize everything in the 6G network architecture—from the physical to application layers—these technologies will drive the change. With the exponential growth in the volume and complexity of data, along with dynamic user requirements, traditional algorithmic approaches are increasingly insufficient. AI/ML technologies step in as powerful enablers, capable of navigating these complexities \cite{wang_road_2023, tataria_6g_2021}.
\\
Over the last decade, the world experienced a shift towards the inclusion of AI technologies in industries such as healthcare and transportation. The integration of AI in 6G will be even more revolutionary. Rather than simply connecting devices, networks will learn, predict behavior, and respond appropriately on an intelligent level \cite{rajoria_brief_2022, viswanathan_communications_2020,FDTPA}. 
At the core of this transformation, there are three fundamental types of ML methods:
\begin{itemize}
    \item \textbf{Supervised Learning:} aims at finding a mapping between inputs and outputs by analyzing labelled training data containing input-output pairs. Typical tasks for supervised learning are classification and regression. In 6G, supervised learning is promising for detection and prediction.
    \item \textbf{Unsupervised Learning:} In unsupervised learning, the training data are unlabelled and contain only inputs, without any indication of the desired outputs. The goal of unsupervised learning is to recognize patterns behind data generation. Unsupervised learning is typically applied to density estimation, clustering, feature extraction, and dimensionality reduction \cite{pennanen_6g_2025}.
    \item \textbf{Reinforcement Learning (RL):} In the case of 6G, RL is useful in teaching networks the most efficient approaches to different strategies through trial and error. It is useful in areas such as automated decision making for power control, intelligent routing, and spectrum management because those fields require instantaneous decisions.
\end{itemize}
Among the several ML techniques, a few stand out as particularly promising for 6G:
\begin{itemize}
    \item \textbf{Deep Learning (DL):} DL is a popular sub-class of ML, which is remarkably powerful in extracting relevant information from large, high-dimensional data sets. DL is well-suited to many types of communication problems, such as network resource management, network access, data traffic prediction, routing, user scheduling, resource allocation, channel estimation, and signal detection. DL is a particularly promising technology for 6G networks since it fits well for diverse 6G-specific challenges at different levels of the network. 
    \item \textbf{Federated Learning (FL):} FL is a distributed ML method. It offers privacy-preserving model training as it does not require raw data to be shared because training is done on the device \cite{yang_federated_2022}. FL can be applied to many common problems in wireless networks, such as resource and spectrum management, resource allocation, user behavior prediction, channel estimation, signal detection, and security/privacy issues \cite{prasad_tera_toward_2025}.
\end{itemize}
As with any other transformative technology, integration of AI/ML with 6G has its challenges. Aside from data privacy, there is model scalability and interoperability across devices which remain unsolved. 6G will not be just another generation of network; embracing AI/ML as a foundational technology will make it a leap toward a real intelligent communication network — one that thinks, learns, and evolves
\cite{jiang_road_2021, cui_overview_2025, letaief_roadmap_2019}.

\section{Open Problems and Insights}
\label{sec:open_problems}

Across today's 6G technologies, there are several technical issues that still need to be addressed for a successful 6G deployment, including security and privacy, spectrum management, energy and sustainability, network complexity and real-time responsiveness:\vspace{0.2cm}

\noindent \textbf{Security and Privacy:} 6G’s pervasive AI and sensing raise serious trust concerns. Experts note that 6G will demand “unprecedented trust” and a zero-trust security architecture from device to core. Ensuring end-to-end data integrity and user privacy (e.g. via encryption, blockchain, or on-device learning) is a major open problem.  In this context, federated learning emerges as a promising solution for privacy-preserving intelligence, enabling a model to be trained on raw data without it being shared, thus protecting data privacy.\vspace{0.2cm}

\noindent \textbf{Spectrum Management:} 6G must handle a vast range of frequencies (from $<$ 1GHz to THz). Coordinating this spectrum (sharing bands, avoiding interference between systems, and aligning internationally) is unsolved. In ultra-dense networks (UDNs), the complexity becomes even more pronounced due to the importance of spectrum reuse and inter-cell interference. Early 6G plans call for enhanced coexistence and spectrum sharing research. \vspace{0.2cm}

\noindent \textbf{Energy and Sustainability:} As noted, minimizing energy per bit is a system-level priority. Trade-offs between performance and power require dynamic optimization. In large-scale 6G deployments, mean-field game (MFG) theory offers a scalable mathematical framework to optimize power control and energy efficiency among a massive number of devices, especially in UDN environments.\vspace{0.2cm}
    
\noindent \textbf{Network Complexity:} The envisioned 6G architecture (combining terrestrial, satellite, aerial, and IoT elements) will be far more complex than 5G. Managing such a network – slicing resources and debugging faults – is a grand challenge.\vspace{0.2cm}

\noindent \textbf{Real-Time Responsiveness:} Pushing latency below a millisecond (for holographic calling, tactile internet) poses challenges across all layers. Hardware processing delay, protocol overhead, and scheduling must be rethought. Fast edge computing and new physical layer waveforms may help, but research on achieving strict end-to-end deadlines remains critical.

\section{Conclusion}
\label{sec:conclusion}

As the deployment of 5G progresses worldwide, researchers have begun exploring the sixth-generation (6G) networks. This paper provides a brief overview of 6G and its key capabilities. It explores essential technologies such as Massive MIMO, THz communication, Reconfigurable Intelligent Surface, and AI/ML, which drive 6G advancements. Additionally, various applications and usage scenarios are highlighted, such as Immersive Communication, Integrated sensing and communication, Integrated AI and communication, ubiquitous connectivity, Hyper-Reliable and Low-Latency Communications and Massive Communications. The paper concludes with an analysis of open problems and insights shaping the path to 6G. Future research directions should tackle critical issues such as scalability, energy efficiency, and privacy in ultra-dense beyond 5G and 6G networks.

\bibliographystyle{IEEEtran}
\bibliography{Review.bib}
\end{document}